# A Sensor Fusion-based GNSS Spoofing Attack Detection Framework for Autonomous Vehicles

**Sagar Dasgupta***
Ph.D. Student
Department of Civil, Construction & Environmental Engineering
The University of Alabama
3014 Cyber Hall, Box 870205
248 Kirkbride Lane, Tuscaloosa, AL 35487
Tel: (864) 624-6210; Email: sdasgupta@crimson.ua.edu

**Mizanur Rahman, Ph.D.**
Assistant Professor
Department of Civil, Construction & Environmental Engineering
The University of Alabama
3015 Cyber Hall, Box 870205
248 Kirkbride Lane, Tuscaloosa, AL 35487
Tuscaloosa, AL 35487
Tel: (205) 348-1717; Email: mizan.rahman@ua.edu

**Mhafuzul Islam, Ph.D.**
3D Computer Vision/Machine Learning Engineer
MicroVision Inc. WA, USA
Email: mdmhafi@clemson.edu
Phone: (864) 986-5446

**Mashrur Chowdhury, Ph.D., P.E., F.ASCE**
Eugene Douglas Mays Chair of Transportation
Glenn Department of Civil Engineering
Clemson University
216 Lowry Hall, Clemson, South Carolina 29634
Tel: (864) 656-3313; Email: mac@clemson.edu

*Corresponding author

Word count: 6,424 words text + 4 tables x 250 words (each) = 7,424 words

Submission date: August 1, 2021

Paper submitted ONLY for presentation at the Transportation Research Board 101st Annual Meeting

**ABSTRACT**
This paper presents a sensor fusion based Global Navigation Satellite System (GNSS) spoofing attack detection framework for autonomous vehicles (AV) that consists of two concurrent strategies: (i) detection of vehicle state using predicted location shift—i.e., distance traveled between two consecutive timestamps—and monitoring of vehicle motion state—i.e., standstill/ in motion; and (ii) detection and classification of turns (i.e., left or right). Data from multiple low-cost in-vehicle sensors (i.e., accelerometer, steering angle sensor, speed sensor, and GNSS) are fused and fed into a recurrent neural network model, which is a long short-term memory (LSTM) network for predicting the location shift, i.e., the distance that an AV travels between two consecutive timestamps. This location shift is then compared with the GNSS-based location shift to detect an attack. We have then combined k-Nearest Neighbors (k-NN) and Dynamic Time Warping (DTW) algorithms to detect and classify left and right turns using data from the steering angle sensor. To prove the efficacy of the sensor fusion-based attack detection framework, attack datasets are created for four unique and sophisticated spoofing attacks—turn-by-turn, overshoot, wrong turn, and stop, using the publicly available real-world Honda Research Institute Driving Dataset (HDD). Our analysis reveals that the sensor fusion-based detection framework successfully detects all four types of spoofing attacks within the required computational latency threshold.

**INTRODUCTION**

Autonomous vehicles (AVs) require accurate, reliable, and continuous real-time localization information from a Global Navigation Satellite System (GNSS) or Global Positioning System (GPS) to perform their autonomous, navigational, security, and safety-critical applications. GNSS is a satellite-based positioning, navigation, and timing (PNT) service. Two levels of services are provided by GPS providers, i) Standard Location Service (SPS) and ii) Precision Location Service (PPS). While the SPS service is free for civil, commercial, and scientific use, more secure and precise PPS service is strictly for government and military use *(1)*. Current autonomous vehicles use SPS, and it is a reasonable assumption that in the future commercial autonomous vehicles will continue to use SPS, as PPS is not available for commercial use. In the rest of the paper, by GNSS, we refer to the SPS or civilian GPS.

The GNSS mostly depends on satellites and radio communications, which are subject to physical degradation of signal strength, natural or unintentional vulnerabilities, and intentional threats *(2)*. Due to the distance between the satellites and the GNSS receiver, the signal strength can be extremely weak at the receiver end. The signal degradation becomes worse in the dense urban setup with high-rise buildings where the signals get reflected, introducing error as well as disrupting continuous GNSS signal availability *(3), (4)*. The GNSS signal can also become disrupted due to natural or unintentional vulnerabilities, such as the absence of GNSS signal because of walls and ceilings in garages and tunnels, and even thick cloud cover in the sky, and signal degradation due to multipath and radio frequency interference *(5), (6)*. Intentional threats can be divided into two types: jamming and spoofing. In intentional jamming, the target AV receiver antenna is flooded by a high-power signal so that the authentic GNSS signal cannot reach the receiver. Among the foregoing, spoofing is the most sophisticated type of attack, as an attacker can temper the authentic GNSS signal structure and transmit inaccurate location information to the target AV *(7)*. A sophisticated spoofing attack refers that a spoofer has the target vehicle's destination, route, and sensor information. In this study, we focus only on sophisticated spoofing attacks as it is difficult to detect this kind of GNSS attacks.

One of the primary purposes of manipulating the GNSS during a spoofing attack is to temper the GNSS signal so that a target AV can be misguided to a wrong destination, which will compromise safety and security AV passengers as well as transportation of goods from an origin to a destination. Generally, an expensive GNSS signal generator is required to perform a spoofing attack. However, the development of low-cost software-defined radios (SDR) has made GNSS spoofing attack easier to carry out *(8)*. Due to the nature of spoofing attacks, there isn't a single attack detection method that can detect all types of GNSS spoofing attacks *(9)*. Therefore, researchers are more concentrating on developing methods that make a spoofing attack more difficult to achieve without being detected.

The techniques of existing spoofing attack detection methods can be classified into four categories: (i) encryption mechanism; (ii) codeless-cross-correlation measures; (iii) signal statistics analysis; and (iv) antenna-based methods *(2)* (these methods are explained further in the related work section). Due to robustness and resilience against wireless signal spoofing attacks and low cost, inertial navigation system (INS) and inertial measurement units (IMU) sensors, such as gyroscope and accelerometer, are used to derive vehicle position and compare with the GNSS data to detect an attack *(10)–(13)*. Moreover, IMU sensor data are also compared with the GNSS output for attack detection *(9)*. However, INS and IMU sensors accumulate errors in determining vehicle position with time which makes these methods unsuitable for developing a robust spoofing attack detection method. With the rapid development of machine learning (ML) and deep learning (DL) algorithms, researchers have also looked at the potency of detecting an attack by analyzing GNSS signals using ML and DL *(14)–(17)*. Although our prior research shows that a GNSS spoofing attack could be detected by predicting distance traveled between two consecutive time stamps *(7)*, a sophisticated spoofing attack, such as a turn-by-turn attack and a wrong trun attack, cannot be detected using this approach.

In this study, a sensor fusion-based GNSS spoofing attack detection framework is developed and examined where low-cost in-vehicle sensor data are used and compared with GNSS data to detect an attack. The framework consists of two strategies. Motivated by the resilience of the inertial sensors against spoofing attacks and robustness of DL algorithms, in the first strategy, low-cost in-vehicle sensors—i.e., speedometer, accelerometer, and steering angle sensor—which are not only available in commercial AV but also in the human-driven vehicles, are fused and feed to a long short-term memory (LSTM) algorithm to predict the distance an AV will travel between two consecutive timestamps. This predicted distance is then compared with the GNSS-derived distance to detect an attack. To make the framework more robust, a turn detection strategy (the second strategy) is also implemented and compared with GNSS to look for any discrepancy related to turning maneuvers. Note that the GNSS spoofing attack detection framework developed in this paper is focused on autonomous vehicles, which navigate through a structured roadway (e.g., surface roadway systems).

The presented framework in this paper is a new addition to the existing GNSS spoofing attack detection approaches because the detection framework can be deployed for the road navigation scenarios using the location level information without directly analyzing the GNSS signal characteristics, whereas most of the existing approaches detect a spoofing attack by analyzing the signal itself. Moreover, although existing INS/IMU-based approaches derive vehicle position or compare a single sensor output with GNSS output, our developed framework fuses multiple sensors and uses an LSTM network to predict a vehicle's location shift, which has not been explored by any researchers so far.

The rest of the paper is arranged as follows. Related work section covers existing spoofing attack detection methods and identifies the research gap. Data preparation section introduces a real-world dataset, which is used to create the attack dataset, and presents the data preparation approach. GNSS spoofing attack models are presented in attack model section. GNSS spoofing attack detection section presents our GNSS spoofing attack detection framework, and the attack detection efficacy against different spoofing attack scenarios and computational complexity are described in evaluation results section. Finally, conclusion section presents the concluding remarks and future research direction.

**RELATED WORK**

As mentioned in the introduction section, existing spoofing attack detection methods use the following approaches for spoofing attack detection: (i) encryption mechanism; (ii) codeless-cross-correlation measures; (iii) signal statistics analysis; and (iv) antenna-based methods *(2)*. The most common GNSS anti-spoofing methods use encryption algorithms to secure the GNSS signals. Although the military commonly uses the encryption mechanism approach to secure GNSS, this is not a cost-effective solution due to its high infrastructural, computational, and management cost. The codeless-cross-correlation measures use the correlation between unknown encrypted GPS L1 P(Y) code signal (L1 is the main GPS carrier signal, and P(Y) code is the precision or secure code) from multiple receivers for detecting a spoofing an attack *(18)–(20)*. With increasing the number of cross-checking receivers, the performance of such an approach also relates to the cost of additional instruments and corresponding signal processing complexity. The GNSS signal statistics analysis-based approaches use different signal features, such as received signal strength (RSS) *(21),* spatial coherency *(22),* pseudo-range measurements, time of advent and signal parameters estimation to detect GNSS spoofing attacks. These approaches require computationally expensive GNSS signal processing algorithms and sophisticated antenna arrays to ensure high spoofing attack detection accuracy.

Besides these approaches, GNSS spoofing attacks can also be detected by comparing vehicle acceleration using IMU, and that derived from GNSS according to *(9)*. Although this approach performs well for the aircraft, it is not suitable for surface vehicles due to the low vehicle dynamics signature. In *(11),* the position derived from IMU sensors (i.e., accelerometer and gyroscope) is compared with GNSS-derived position for spoofing attack detection. INS has also been used to monitor the position of a vehicle for detecting GNSS spoofing attacks *(10), (12)*. INS device uses the sensor data from the gyroscope and accelerometer and calculates the position, orientation and velocity using dead reckoning without any input from GNSS. However, INS derived position is less accurate as the measurements from inertial sensors accumulate bias as well as scale factor and non-orthogonality errors that accumulate over time. Besides, GNSS signals from multiple antennas are also used for spoofing attack detection *(23)*.

In addition to above mentioned approaches, the development of ML and DL algorithms has recently increased for spoofing attack detection. In (14), a Multi-Layer Perception (MLP), a Complex Convolution Neural Networks (CNN), and a Simple CNN are used to detect spoofed signals. The paper presented the potency of using deep neural networks for spoofing signal detection. In *(16),* the authors present a decision fusion with the K-out-of-N decision rule-based method using the wavelet transformation coefficients, which serves as a general identification scheme to detect GNSS spoofing attacks. In *(15),* a Support Vector Machine (SVM) has been used for state estimation and detect an attack on unmanned aerial vehicles based on it. In *(17),* spoofing attacks are detected using K-Nearest Neighbor (KNN) and naive Bayesian classifier and the early-late phase that capture phase delay between the line of sight (LOS) and reflected signal, delta, and signal level features of GPS. However, these ML and DL algorithms are used to detect an attack in the signal level, and no work has been done to predict the distance the vehicle can travel within a timeframe and detecting an attack based on that. Only INS-based spoofing attack detection approaches, where we compare INS-based vehicle's position with the GNSS-based position, are closely related to our study. However, position information from INS sensors is not reliable due to the error propagation of inertial sensors over time. Thus, none of the approaches used the location domain information to detect the spoofing attack.

Our study differs from the previous studies as we have used the speedometer, steering wheel angle data, and accelerometer data to predict the location shift/distance traveled by the next timestamp using an artificial recurrent neural network (RNN) architecture LSTM. The predicted location shift by the AV is then compared with the GNSS-based location shift to detect any spoofing attack. Moreover, a turn detection strategy is used to further detect more sophisticated attacks. Our study is novel because none of the existing research uses predicted location shift and turn detection techniques for developing sophisticated spoofing attack detection framework utilizing GNSS information (latitude, longitude and turn type) and in-vehicle sensor data without analyzing the GNSS physical signal characteristics.

**DATA PREPARATION**

The HDD (24) is used in this study to develop and evaluate the GNSS attack detection framework. The HDD contains data from the camera, LiDAR, GNSS, inertial measurement unit (IMU), and controller area network (CAN) of a conventional vehicle, and it is collected from suburban and urban roadways as well as highways within the San Francisco Bay Area. As AVs are equipped with cameras, IMU, GNSS, the HDD is suitable for developing technologies for AVs. **Figure 1** shows a sample route from the HDD.

In HDD, the GNSS signals are recorded at 120Hz using a GeneSys Eletronik GmbH Automotive Dynamic Motion Analyzer with a Differential Global Positioning System (DGPS). The acceleration (m/s$^2$), steering wheel angle (deg), rotational speed of the steering wheel (deg/s), vehicle speed (ft/s), brake pressure (kPa), and yaw rate (deg/s) are collected from different sensors and recorded from each vehicle's CAN bus at 100Hz. For our detection framework development and evaluation, the latitude and longitude, relative accelerator pedal position (%), which represents the acceleration of an AV, steering wheel angle, and speed data are extracted from the HDD. We have then synchronized the extracted HDD by keeping GNSS UNIX timestamp as a reference by interpolating between two closest observations. We have also calculated the perceived location shift, i.e., the distance traveled between two consecutive timestamps, with data from GNSS using the Haversine formula (see **Equation 1**) (25):

$$d = 2r \sin^{-1}\left(\sqrt{\sin^2\left(\frac{\varphi_2 - \varphi_1}{2}\right) + \cos(\varphi_1)\cos(\varphi_2)\sin^2\left(\frac{\psi_2 - \psi_1}{2}\right)}\right) \quad (1)$$

where d is the distance in meter between two points on the Earth's surface; r is the Earth's radius (6378 km); $\varphi_1$ and $\varphi_2$ are the latitudes in radians; $\psi_1$ and $\psi_2$ are the longitudes in radians of two consecutive time stamps.

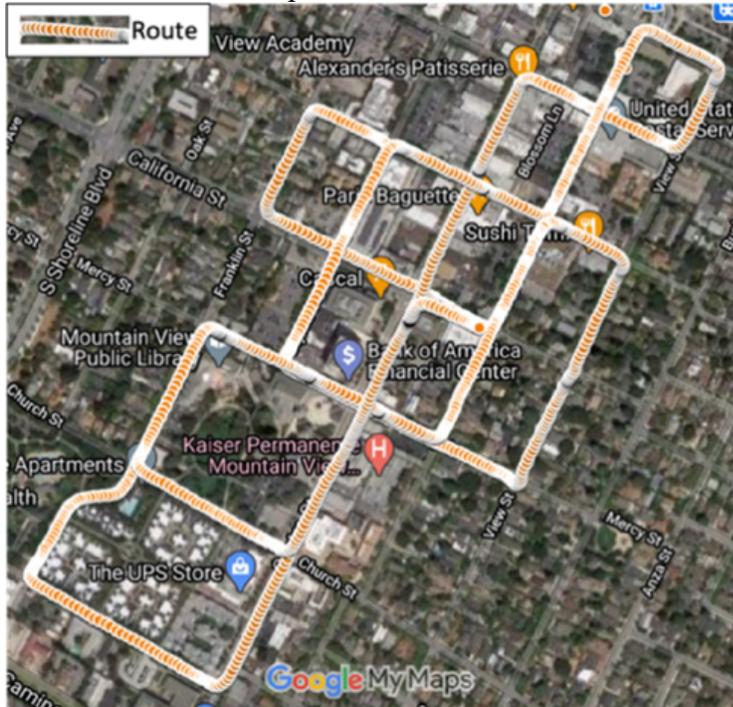

**Figure 1 An example of GNSS traces from the HDD.**

**ATTACK MODELS**

The developed GNSS spoofing attack detection framework is evaluated against four sophisticated spoofing attack scenarios, which are: (a) turn-by-turn attack; (b) overshoot attack; (c) wrong turn attack; and (d) stop attack. A turn-by-turn attack is a sophisticated attack because the spoofer has an AV's route and destination information. In this type of spoofing attack (26), the spoofer takes over an target AV's GNSS and changes the AV's current location resulting in a location shift between the AV's location before and after the attack. Due to a change in AV's current location, the navigation application creates a new route to reach the destination, and the AV believes in the spoofed location, follows the newly created wrong route, and ends up in a wrong, possibly unsafe location instead of a desired destination. The spoofer also tries to keep realistic values of the location shift, AV's speed, and distance between actual and spoofed routes to make it more believable.

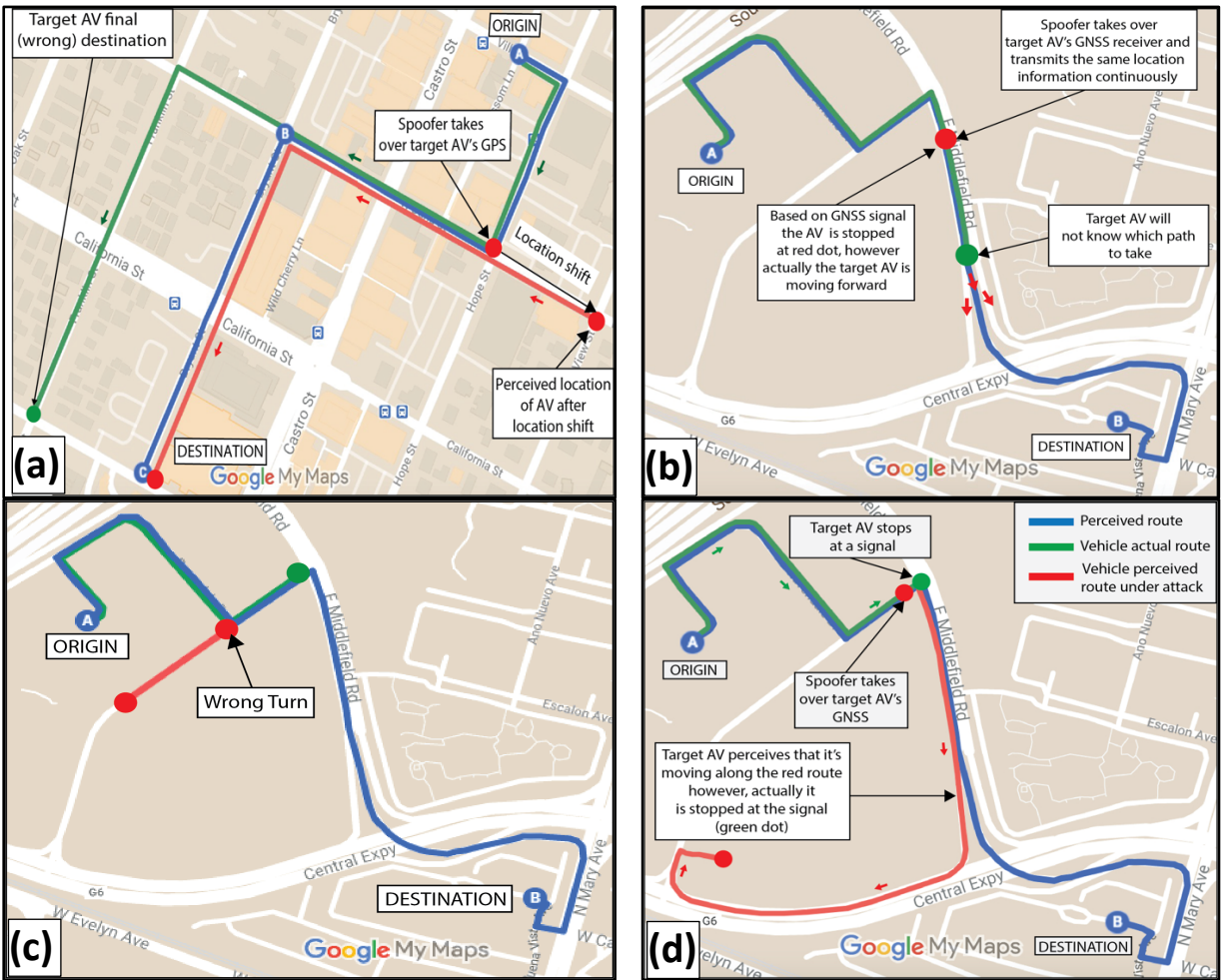

**Figure 2 GNSS spoofing attack models: (a) example of turn-by-turn; (b) example of overshoot; (c) example of a wrong turn and (d) example of a stop**

**Figure 2(a)** illustrates a turn-by-turn attack in which the correct route from the origin to destination is shown in blue, the AV's ground truth route is shown in green, and the AV's perceived route, which matches with the original route turn-by-turn, is shown in red. Thus, compromising the GNSS, a spoofer creates a wrong route matching for the new route's number of turns and guides the vehicle to a wrong destination by compromising the AV's GNSS.

**Figure 2(b)** shows an overshoot attack (27). After taking over the AV's GNSS, the spoofer keeps sending the same location information. As a result, based on the GNSS output, the AV perceives that it is in a standstill state (shown as stopped at the red dot), but the AV is moving forward in reality. When a road split (at the green dot) or intersection occurs, the AV will be unable to identify the path to proceed.

During a wrong turn attack, the spoofer takes over the target AV's GNSS receiver just before a turn. While the target AV takes a right turn, as shown in **Figure 2(c)**, the spoofer tempers the GNSS signal, and the target AV perceives that it is taking a left turn. Similarly, if the AV takes a left turn, the GNSS will show a right turn. This will lead to rerouting of the attacked AV, and the target AV vehicle will end up at the wrong destination.

A stop attack (27) (see **Figure 2(d)**) is the opposite of an overshoot attack. The spoofer takes over the receiver GNSS when the AV is stopped at a stop sign (green dot) or due to traffic and then transmits a synthetic signal so that the AV perceives that it is moving along the road (shown as red route). We have created ten attack scenarios for each of these four unique attack types using data from the HDD to mimic the actual spoofing scenarios

**GNSS SPOOFING ATTACK DETECTION FRAMEWORK**

We have developed a robust GNSS spoofing attack detection framework (see **Figure 3**), and this framework involves two concurrent strategies in which data from in-vehicle low-cost sensors—i.e., GNSS, accelerometer, steering wheel angle sensor, and speedometer—are fused to provide a unified and robust GNSS spoofing attack detection strategy. These two strategies include: (i) detection of a vehicle's state using predicted location shift and vehicle motion state information, and (ii) detection of turning maneuvers (right and left turns).

In the first strategy, our primary objective is to develop a vehicle state prediction model so that the model can predict the subject vehicle's state information (i.e., distance traveled/location shift) by fusing data from multiple in-vehicle sensors (i.e., speedometer, accelerometer, steering angle sensor). For every timestamp, attack free speed, acceleration, location shift in previous timestamps, and steering angle data of the AV are fed to train a deep recurrent neural network model, which is LSTM, that will predict the location shift between two consecutive timestamps. This model can predict the location shift considering the long-term dependencies by storing the temporal dependency of the time-series data in the recurrent hidden layer's memory blocks. Along with checking the vehicle location shift, this strategy also continuously checks if the vehicle is in motion or in a standstill state using the speedometer output. Because the predicted location shift data alone cannot reliably measure an AV's motion state. If the speedometer output is less than the speedometer sensor's maximum error value, then the vehicle is in a standstill state, and if it is greater than the error threshold, the vehicle is in motion.

According to the second strategy, steering angle data can be used to recognize different types of turns (left or right). For example, steering angle sensor or gyroscope output can provide turn maneuvering data. In this study, we use steering angle sensor output instead of gyroscope as the steering angle data provide better vehicle maneuvering data and are available in the HDD. A vehicle's turn can be divided into three categories: left-turns, right-turns, and U-turns. In this paper, we have only concentrated on detecting left and right turns, which are the most common types of turns. Because of variability in driving maneuvers with the different roadway turning curvatures, the length of the duration of the maneuvering data of the steering angle sensor may differ. Therefore, in recognition of vehicle maneuvers, different types of turning maneuver data for each type of left and right turns from the HDD dataset are used to train the dynamic time warping (DTW) algorithm, which can measure the similarity between two time series data to recognize a turn pattern and a k-NN classifier algorithm is used to classify different turns. The detection system continuously compares the output from the inertial sensors with turning information from GNSS to detect and classify turning maneuvers. There is a possibility that an AV is at a standstill state, but the steering wheel is rotating for adjustment; in such a case, the steering wheel data can resemble a turn pattern, but no turning maneuver occurs in reality. To address such scenario, the turn detection strategy also takes input from the speedometer. If a turn signal is detected along with vehicle speed higher than the maximum possible error, our strategy will detect a turn; otherwise, no turn will be detected. It is a common scenario when an AV will be at a parking lot and try to park itself. Overall, any GNSS attack can be detected based on these two strategies—i.e., (i) detection of a vehicle's state using predicted location shift and vehicle motion state information, and (ii) detection of turning maneuvers (right and left turns).

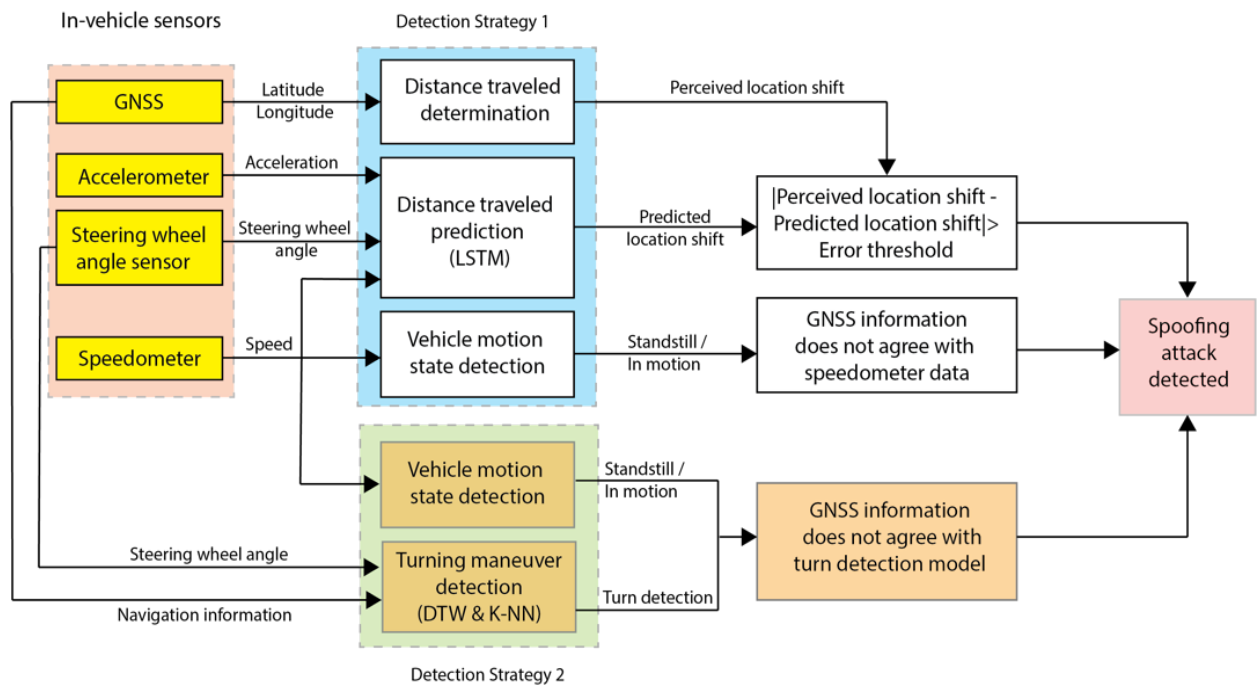

**Figure 3** Sensor fusion based GNSS spoofing attack detection framework

**Development of Detection Strategy 1**

The first strategy incorporates two different types of vehicle state information: (a) predicted and perceived location shift between two consecutive timestamps; and (b) vehicle motion state detection (standstill/in motion). The location shift by an AV between two consecutive timestamps is predicted using a 2-stacked LSTM (28) with 128 and 64 neurons in the first and second hidden layers, respectively. The training and validation data includes the previous location shifts between two consecutive timestamps, acceleration, steering wheel angle, and speed. The output is the location shift between the current timestamp and the immediate future timestamp. In this study, the data generation frequency is 120Hz or 0.00833s, which is the time difference between two consecutive timestamps (24). The dataset is splited into training with 241,990 observations and validation with 103,709 observations. Before feeding the sensor output to the LSTM training, the input features are normalized between 0 and 1. The LSTM hyperparameters, i.e., number of neurons, number of epochs, batch size and learning rate, are selected by a trial-and-error approach

(8) as it is a time series-based prediction model. The Mean Absolute Error (MAE) metric is used as the loss function for evaluating overfitting and underfitting while training the model. To evaluate the goodness-of-fit of the LSTM model the loss profiles using optimal hyperparameters are shown in **Figure 4** with both the training and validation datasets. A comparison of the MAE of these two datasets indicates a good fit of the prediction model with the optimal hyperparameters. The values of hyperparameters and name of the optimizer are listed in **Table 1**. After testing the prediction model, we found that the Root Mean Square Error (RMSE) for the predicted location shift is 0.0278 m, and the maximum absolute error is 0.0446 m.

A threshold is established by adding the prediction model maximum absolute error and GNSS positioning error as given in (2). Based on the GNSS device model used for collecting the HDD, the max positioning error is 0.1 m (4). Therefore, the error threshold value is 0.1446 m. If the difference between the perceived location shift using GNSS and the predicted location shift is greater than the error threshold, an attack will be detected.

$$Error\ Threshold = Prediction\ Model\ Maximum\ Absolute\ Error$$
$$+ Positioning\ Error\ of\ the\ GNSS \qquad (2)$$

Along with checking the vehicle location shift, if the speedometer output is less than the speedometer sensor's maximum error value, then the vehicle is in a standstill state, and if it is greater than the error threshold, the vehicle is in motion.

**TABLE 1 LSTM Model Hyperparameters**

| Hyperparameters and Optimizer | Value |
| --- | --- |
| Number of neurons (1st layer) | 128 |
| Number of neurons (2nd layer) | 64 |
| Number of epochs | 500 |
| Batch size | 50 |
| Learning rate | 0.01 |
| Optimizer | Adam |

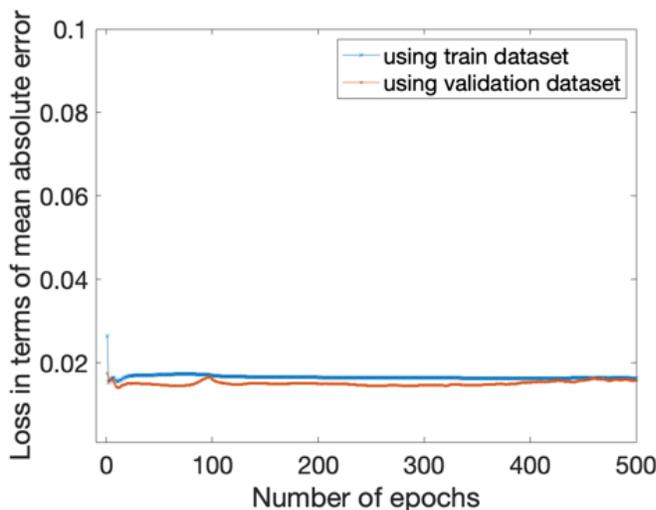

**Figure 4** Comparison of Mean Absolute Error (MAE) (loss) profiles with the optimal parameter set

**Development of Detection Strategy 2**

According to the second strategy, turn detection, a vehicle turning maneuver is detected and classified by comparing the steering angle sensor output and the GNSS output. If a turn is detected using steering angle data, but GNSS shows no turn, then an attack will be detected. Moreover, if the steering angle change represents a right turn and GNSS detects a left turn, an attack will be detected. The AV steering angle readings create unique shapes for left and right turns (as shown in the top left corner of **Figure 5**). For example, for taking a right turn, an AV first turns the steering wheel right to get into the road and then turns the steering wheel to the left to straighten the vehicle creating a unique shaped curve (see **Figure 5**). However, due to varying steering behavior and road geometry, the length of the duration of the maneuvering data for a turn are not uniform.

A k-Nearest Neighbors (k-NN) clustering algorithm is combined with a dynamic time warping (DTW) algorithm for developing left and right turn detection. The classification model and steering wheel angle data are used as model input. The DTW compares the patterns and measures the similarity between two different time-series data of the different number of observations *(29)*. The DTW iteratively warps the time axis to align two input time series and searches for an optimal match; it then calculates the warp path distance, which is the cumulative distance between each pair of observations. The path with minimum total cost represents the DTW distance between two different time series data as shown in **Equation 3**:

$$DTW(T,S) = \underset{w = w_1, w_2 \ldots, w_k, \ldots, w_K}{argmin} \sqrt{\sum_{k=1, w_k=(i,j)}^{K} (t_i - s_j)^2} \qquad (3)$$

where T and S are ground truth and training steering angle data respectively; $w$ represents a warping path; $t_i$ is the ith observation of time series T; and $s_j$ is the jth observation of the time series S. We have used FastDTW (30) algorithm to reduce the computational time and satisfy the real-time detection requirement.

k-NN is a widely used classifier. The k-NN algorithm assigns a common class among its k nearest neighbors based on a distance metric. In this study, DTW is used as the distance metric for k-NN. The steering angle reading of right and left turns from the HDD is used for training and testing the k-NN model. The training dataset contains 19 right turns and 13 left turns, and the testing dataset contains 7 right turns and 6 left turns (see **Table 2**). The sample left, and right turn locations using black and purple lines are shown in **Figure 5**. The combined k-NN—DTW turn detection model can classify left, and right turns with an accuracy of 100% along with precision value, recall value, and $F_1$-score of 1. Note that our training dataset contains different patterns of left and right turns.

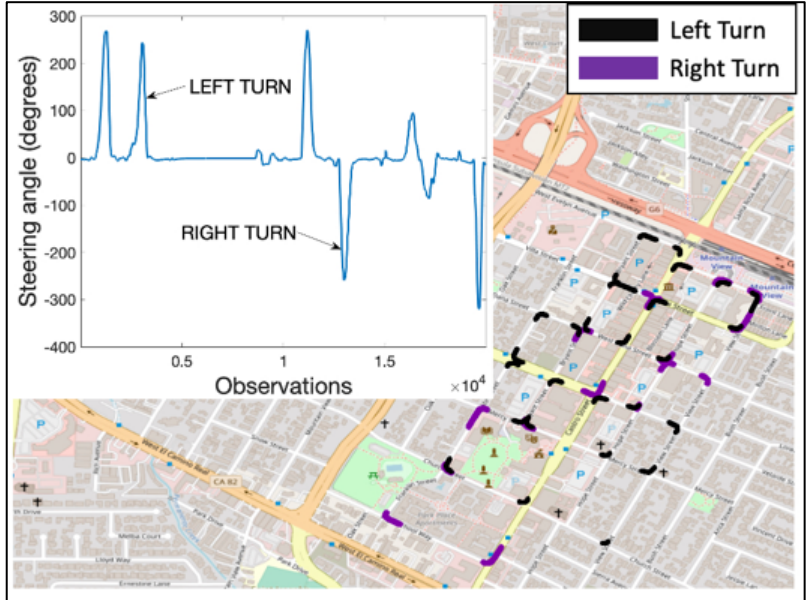

**Figure 5 Patterns of right and left turning maneuvers**

**TABLE 2 Data used for k-NN-DTW Model Training**

| Turn type | Dataset | Number of turns | Number of observations |
|---|---|---|---|
| Right | training | 19 | 15969 |
| | testing | 7 | 8209 |
| Left | training | 13 | 12974 |
| | testing | 6 | 6706 |

**EVALUATION RESULTS**

To prove the efficacy of our detection framework, we have evaluated the detection framework using different spoofing scenarios (see "Attack Models" section and **Table 3**) and validated the performance of our framework by comparing with the ground truth attack data, i.e., baseline, for ten attack scenarios for each of the following attack types: turn-by-turn, overshoot, wrong turn and stop. Each attack dataset contains data before and after the spoofing attack. After the spoofing attack, the dataset represents compromised GNSS latitude and longitude data along with other in-vehicle sensors data (acceleration pedal position, speed) based on spoofing attack scenarios. These datasets represent actual attack scenarios as it contains real-world driving data after and before an attack.

**Table 3** presents the attack evaluation scenarios and corresponding attack strategies for all four types of spoofing attacks considered in this study. The turn-by-turn attack is detected using the location shift prediction strategy (strategy 1). Both strategies 1 and 2 are simultaneously used to detect the overshoot and stop the attack. Suppose the spoofer manages to update the location within the error threshold. Even in that case, the attack is detected due to the combination of strategies utilized here, which continuously check the vehicle motion state in both strategies. For a wrong turn type of attack, the second strategy is used where the turn-based on the GNSS information is compared with the steering angle sensor data.

**TABLE 3 Attack Evaluation Types, Number of Scenarios and Corresponding Detection Strategies**

| Attack type | Number of scenarios | Detection Strategies | |
|---|---|---|---|
| | | Strategy 1 | Strategy 2 |
| Turn-by-turn | 10 | ✓ | NA |
| Overshoot | 10 | ✓ | ✓ |
| Wrong turn | 10 | NA | ✓ |
| Stop | 10 | ✓ | ✓ |
| Note: ✓ - Required, NA - Not applicable | | | |

A turn-by-turn type GNSS spoofing attack detection is presented in **Figure 6**, where the observation number is shown on the x-axis, and the y axis represents the difference between perceived and predicted location shifts. In this scenario, the location shift for the uncompromised GNSS case in shown in **Figure 6(a)** where the location shifts never cross the threshold (see "Development of Detection Strategy" subsection and **Equation 2**). When an attack is generated, the difference between the perceived and predicted location shift (**Figure 6(b)**) crosses the threshold, thus the attack is detected.

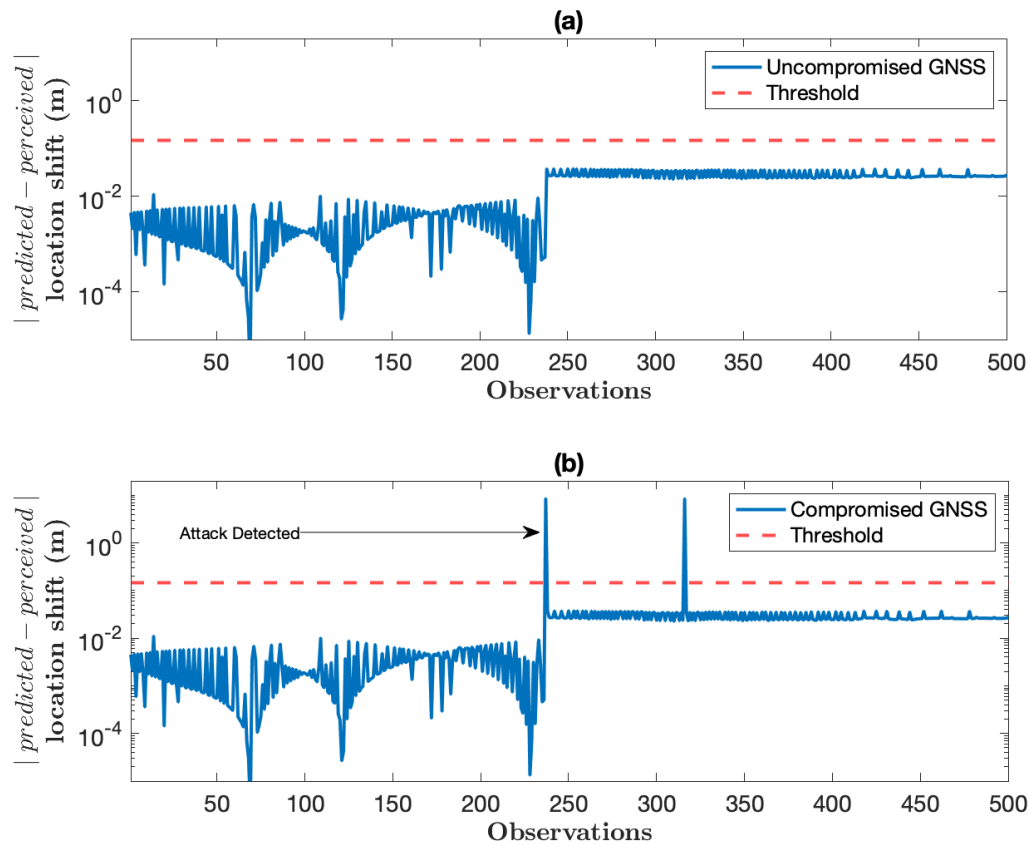

**Figure 6 An example GNSS spoofing scenario: (a) location shift for an uncompromised GNSS observations; and (b) location shift for a compromised GNSS observations**

**Figure 7** presents profiles for the absolute difference between the perceived location shift and predicted location shift for ten different turn-by-turn type spoofing attack scenarios (AS). Note that, the absolute difference between the perceived location shift and predicted location shift is plotted on log scale in **Figure 7** as the difference is negligible. At the point of beginning of the spoofing attack (as shown using different markers), the difference between perceived and predicted location shift is higher than the error threshold value; thus, it detects the attack. It is worth mentioning that no false attack is detected using our detection framework (see **Figure 7**).

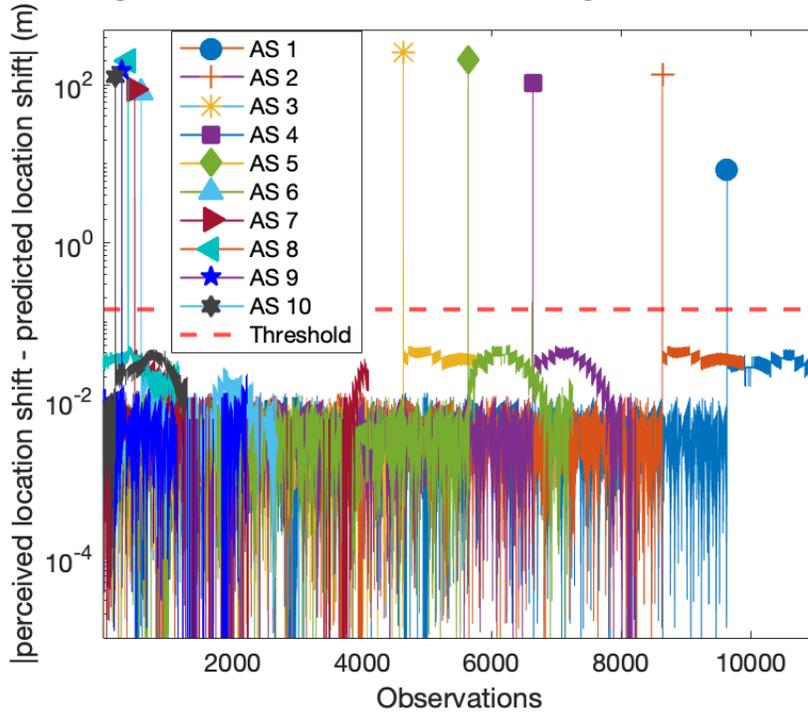

**Figure 7 Attack detection for ten turn-by-turn attack scenarios (AS)**

The efficacy of the turn detection and classification model using the K-NN—DTW combined method is shown in **Table 4**. As the calculated precision and recall is 1, all turns are correctly detected and classified, and there is no false detection and classification occurred. The $F_1$ score is also 1 (100% effective in turn detection and classification), which proves the efficacy of the turn detection strategy. Overall, the results reveals that our detection framework can detect all four attack types successfully in each of all 10 scenarios.

**TABLE 4 k-NN and DTW Testing Output**

| Perceived turn type via GNSS | Number of attack scenarios | Detected turn types using k-NN and DTW | Baseline/actual turn type (ground truth data) | Accuracy/Precision/Recall/ $F_1$ Score |
|---|---|---|---|---|
| Left Turn | 5 | Right turns | Right turns | |
| Right Turn | 5 | Left turns | Left turns | 1.00 |
| Left or Right turns | 20 | No turns | No turns | |

The average computational latency for our first strategy, i.e., location shift prediction strategy— is 0.691μs for each observation, which is less than the GNSS data generation frequency (i.e., 120Hz or 0.0083s or 8300μs). In our second strategy, we have resampled the steering angle sensor data to 5Hz from 120Hz because our experiments showed that 5Hz sampling rate reduces the computational time while preserving the observational (or sensing) integrity of the right and left turns. The k-NN—DTW model takes 0.08s on average to detect a turn, which is less than the data sampling frequency (i.e., 5Hz or 0.2s). Note that, for the computational time presented in this study applies to a workstation, equipped with dual Intel Xeon Gold 5215 2.5GHz processor with 128GB DDR4 2666MHz RDIMM ECC RAM memories, used to run our experiments.

**CONCLUSIONS**

A robust GNSS spoofing attack detection framework is presented in this paper. Data from low-cost in-vehicle sensors are used for detecting sophisticated GNSS spoofing attacks. The framework presented in this paper is unique compared to existing approaches in two ways. First, a deep learning algorithm is used to predict, whereas existing approaches use deep learning to analyze GNSS signal to detect an attack. Second, although existing INS/IMU-based approaches derive vehicle position or compare a single sensor output with GNSS data for detecting cyberattacks on GNSS, our framework fuses multiple sensors to predict a vehicle's location shift and detect turn, which has not been explored by any researchers so far. Instead, in our first strategy, we have used data from multiple sensors –i.e., speedometer, steering angle, and accelerometer— to predict the location shift/distance traveled by the next timestamp using an artificial recurrent neural network (RNN) architecture LSTM. The predicted location shift by the AV is then compared with the GNSS-based location shift to detect any spoofing attack. Moreover, a turn detection strategy, which is our second strategy, is used for classifying turns to further detect more sophisticated attacks. The combination of two strategies allows the framework to detect the most sophisticated spoofing attacks where a spoofer has the capability of tempering the target vehicle's destination, route, and sensor information (i.e., the motion state of an AV). The framework presented is also validated against four different attack types considered in this study. Our attack detection framework is able to detect different types of attacks with a high degree of success. Further research can be performed to evaluate and validate the framework through real-world experiments.

**ACKNOWLEDGMENT**

This material is based on a study partially supported by the National Science Foundation under Grant No. 2104999. Any opinions, findings, and conclusions or recommendations expressed in this material are those of the author(s) and do not necessarily reflect the views of the National Science Foundation, and the U.S. Government assumes no liability for the contents or use thereof.

**AUTHOR CONTRIBUTIONS**

The authors confirm contribution to the paper as follows: study conception and design: S. Dasgupta, M. Rahman, M. Islam and M. Chowdhury; data collection: S. Dasgupta, M. Rahman, and M. Islam; interpretation of results: S. Dasgupta, M. Rahman, M. Islam and M. Chowdhury; draft manuscript preparation: S. Dasgupta, M. Rahman, M. Islam and M. Chowdhury. All authors reviewed the results and approved the final version of the manuscript.